\documentclass[manuscript,screen]{acmart}
\usepackage[most]{tcolorbox}
\usepackage{hyperref}
\usepackage{graphicx}    
\usepackage{array} 
\usepackage{booktabs} 
\usepackage{multirow} 
\usepackage{siunitx}
\usepackage{comment}
\usepackage{setspace}
\usepackage{titlesec}
\usepackage{textgreek}
\usepackage[table]{xcolor}
\usepackage{booktabs}
\usepackage{multirow}
\usepackage{tcolorbox}
\usepackage{pifont}
\usepackage{enumitem}

\titlespacing*{\subsubsection}{0pt}{10pt}{4pt}

\AtBeginDocument{%
  }

\setcopyright{acmlicensed}
\copyrightyear{2026}
\acmYear{2026}
\acmDOI{XXXXXXX.XXXXXXX}

\acmConference[Conference acronym 'XX]{Make sure to enter the correct
  conference title from your rights confirmation email}{June 03--05,
  2026}{Woodstock, NY}

\acmISBN{978-1-4503-XXXX-X/2018/06}

\begin{document}
  
\title{Behaviour Driven Development Scenario Generation with Large Language Models}

\author{Amila Rathnayake}
\email{amila.rathnayake@student.rmit.edu.au}
\affiliation{%
  \institution{School of Computing Technologies, RMIT University}
  \country{Australia}}

\author{Mojtaba Shahin}
\email{mojtaba.shahin@rmit.edu.au}
\affiliation{%
  \institution{School of Computing Technologies, RMIT University}
  \country{Australia}} 

  \author{Golnoush Abaei}
  \email{golnoush.abaei@rmit.edu.au}
\affiliation{%
  \institution{School of Computing Technologies,  RMIT University}
  \country{Australia}}

\begin{abstract}
This paper presents an evaluation of three LLMs, GPT-4, Claude 3, and Gemini, for automated Behaviour-Driven Development (BDD) scenarios generation. To support this evaluation, we constructed a dataset of 500 user stories, requirement descriptions, and their corresponding BDD scenarios, drawn from four proprietary software products. We assessed the quality of BDD scenarios generated by LLMs using a multidimensional evaluation framework encompassing text and semantic similarity metrics, LLM-based evaluation, and human expert assessment. Our findings reveal that although GPT-4 achieves higher scores in text and semantic similarity metrics, Claude 3 produces scenarios rated highest by both human experts and LLM-based evaluators. LLM-based evaluators, particularly DeepSeek, show a stronger correlation with human judgment than with text similarity and semantic similarity metrics. The effectiveness of prompting techniques is model-specific: GPT-4 performs best with zero-shot, Claude 3 benefits from chain-of-thought reasoning, and Gemini achieves optimal results with few-shot examples. Input quality determines the effectiveness of BDD scenario generation: detailed requirement descriptions alone yield high-quality scenarios, whereas user stories alone yield low-quality scenarios. Our experiments indicate that setting temperature to 0 and top\_p to 1.0 produced the highest-quality BDD scenarios across all models.

\end{abstract}




\maketitle

\section{Introduction}

In the fast-changing world of software development, effective testing methodologies have become critical to ensuring software quality. Modern software systems are characterised by unprecedented complexity, with applications spanning multiple platforms, integrating diverse technologies, and serving millions of users with varying requirements \cite{wang2020automatic, escalona2011overview}. This complexity, coupled with accelerated development cycles driven by agile methodologies and continuous delivery practices, has created significant challenges for traditional testing approaches that rely heavily on manual processes.

The software testing industry faces a convergence of pressures that demand innovative solutions. Development teams are expected to deliver high-quality software faster than ever before, while ensuring comprehensive test coverage across an expanding range of scenarios, edge cases, and user workflows. Traditional manual testing approaches cannot scale effectively to meet these demands due to resource constraints, time limitations, and variability in human-generated test scenarios \cite{garousi2016and, ricca2025next, karpurapu2024comprehensive}. This scalability crisis has led to increased interest in automated testing solutions that can maintain quality standards while reducing the time and expertise required for comprehensive test coverage.

Addressing these scalability challenges requires not only automation but also methodologies that can bridge the communication gap between business requirements and technical implementation  \cite{binamungu2018maintaining, binamungu2023behaviour}. Behaviour-Driven Development (BDD) is an approach in agile environments that fundamentally changes how software requirements are communicated, understood, and validated \cite{binamungu2023behaviour}. Unlike traditional testing approaches, which sometimes create a divide between technical and business stakeholders, BDD establishes a collaborative framework that bridges this gap through a shared, natural-language approach to requirements specification \cite{irshad2021adapting,binamungu2023behaviour}. The methodology focuses on three key activities: discovery (structured conversations between stakeholders), formulation (translation of requirements into Gherkin Given/When/Then scenarios), and automation (conversion of scenarios into executable test cases) \cite{zameni2023bdd}. These scenarios serve as both living documentation and executable test specifications, making BDD valuable for ensuring software quality while maintaining stakeholder alignment \cite{zameni2023bdd}.

Despite its benefits, BDD faces significant challenges that limit its adoption and effectiveness. The manual creation of comprehensive BDD scenarios is time-consuming \cite{karpurapu2024comprehensive}. This process creates bottlenecks in agile development environments, where rapid iteration is essential. Additionally, scenario quality varies significantly based on the author's expertise and experience, leading to inconsistent test coverage and potential gaps in edge case identification.

Large Language Models (LLMs) have demonstrated outstanding capabilities in natural language interpretation and structured text generation, with proven success across various software engineering tasks, including requirements classification, code generation, test case generation, and documentation creation \cite{hou2024large,fan2023large,wang2024software}. Their ability to understand requirements expressed in natural language and generate well-structured, contextually appropriate output makes them potentially suitable for automating BDD scenario creation. However, current research in this area remains limited. While prior studies have shown that models such as GPT-3.5 and GPT-4 can generate syntactically correct BDD scenarios \cite{karpurapu2024comprehensive}, there is insufficient systematic evaluation of their effectiveness across multiple quality dimensions. Critical gaps exist in understanding how different LLMs compare for BDD scenario generation, which evaluation methods are most suitable for assessing generated scenario quality, and how automated metrics correlate with human expert assessment.

This research addresses these gaps by conducting a comprehensive evaluation of three representative LLMs (GPT-4, Claude 3, and Gemini) for BDD scenario generation. Given the absence of a publicly available dataset of BDD scenarios, we first constructed a dataset comprising 500 user stories, requirement descriptions, and their corresponding BDD scenarios collected from four industrial software products within a software company. We then evaluated the BDD scenarios generated by LLMs using a wide range of metrics, including text and semantic similarity metrics, LLM-based evaluation, and human assessments. Our findings show that all evaluated models produce promising BDD scenarios. Notably, when employed as an LLM-based evaluator, DeepSeek demonstrates a strong positive correlation with human judgments. Next, we investigated the impact of different prompting techniques on the quality of BDD scenarios, observing mixed effects across models. We also analysed how different input configurations (user story plus requirement description, user story only, and requirement description only) affect the quality of BDD scenarios. The findings reveal that detailed requirement descriptions, whether provided alone or in combination with user stories, produce higher-quality BDD scenarios than user stories alone. Finally, an investigation of different LLM parameter settings (temperature and top\_p) shows that a temperature of 0 and a top\_p of 1 consistently produce the highest-quality BDD scenarios.

Our key contributions include:
\begin{itemize}
\item We construct the first dataset comprising 500 user stories, requirement descriptions, and their corresponding BDD scenarios.
\item To our knowledge, this work presents the first comprehensive study that evaluates the ability of LLMs to generate BDD scenarios from multiple perspectives.
\item We offer recommendations and practical guidance to improve the efficiency and quality of LLM-based BDD scenario generation.
\item We publicly release our dataset and source code to facilitate reproducibility and support future research and practical applications \cite{rathnayake2026llmbdd}.
\end{itemize}

The rest of the paper is organised as follows: Section \ref{sec:example} presents a motivating example, followed by background on BDD and related work in Section \ref{sec:background}. We outline our methodology in Section \ref{sec:method}. Section \ref{sec:findings} reports the empirical findings, while Section \ref{sec:discussion} discusses these findings and highlights their implications for research and practice. Section \ref{sec:threats} examines potential threats to validity and the corresponding mitigation strategies. Finally, Section \ref{sec:coclusion} concludes the paper and introduces future directions.

\section{Motivating Example} \label{sec:example}

Consider that a team working on a digital asset management platform at IntelligenceBank\footnote{\href{https://www.intelligencebank.com/}{https://www.intelligencebank.com}} wants to implement a feature that allows users to apply complex permissions to shared files. The Quality Assurance (QA) lead, Sarah, receives a user story describing the new permissions feature:
\textit{“As a content manager, I want to apply time-bound permissions to shared files so that external users can only access files during a specified timeframe”}. 

\textbf {Current Scenario:} Following BDD practices, Sarah takes the steps below to create a BDD scenario.
\begin{enumerate} [leftmargin=4ex]
\item Manually analyse the user story and description from Jira, then identify all the information to prepare a test scenario (temporary access, expiration behaviour, permission conflicts, etc.).
\item Create a Gherkin scenario to cover all the requirements.
\item Ensure scenario meets both technical accuracy and business requirements.
\item Review a scenario with developers and product owners, making revisions through multiple feedback cycles.
\end{enumerate}

For this single feature, Sarah spends approximately hours creating a comprehensive BDD scenario. The process delays testing, creates a bottleneck in the development cycle, and sometimes results in missing edge cases due to time constraints.

\textbf {With LLM-based BDD scenario generation:}
Using LLMs, Sarah inputs the user story and its requirement description into an LLM:

\begin{enumerate} [leftmargin=4ex]
\item The LLM analyses the user story and description, then automatically generates a comprehensive BDD scenario that covers core functionality, edge cases, and potential error conditions.
\item The scenario follows proper Gherkin syntax and BDD best practices.
\item Sarah can quickly review and make minor adjustments rather than creating a scenario from scratch.
\end{enumerate}

The entire process takes less time than the manual process. The result is not just time savings but significantly improved test coverage. The LLM identifies several edge case details Sarah might have missed, such as handling permission conflicts when files are moved between folders with different access settings, or timezone considerations for expiration dates.
This example demonstrates how our research directly addresses real challenges in software testing by combining the efficiency of automation with LLMs, ultimately leading to better software quality and faster development cycles.

\section{Background and Related Work}\label{sec:background}

This section reviews the relevant literature across several areas to establish the foundation for this research. We first provide a background on BDD, followed by an examination of the existing literature on challenges in BDD implementation. We then outline traditional approaches to BDD scenarios and the use of LLMs in software testing.

\subsection {Background: BDD}

BDD evolved from Test-Driven Development (TDD) as a methodology aimed at bridging the communication gap between technical and non-technical stakeholders in software development \cite{binamungu2023behaviour}. According to Zameni et al. \cite{zameni2023bdd}, BDD establishes a collaborative framework on three key activities: discovery, formulation, and automation.
In the discovery phase, developers, quality assurance analysts, and business stakeholders engage in structured conversations to explore the required software behaviours. These conversations typically use concrete examples to establish a shared understanding of the system functionality. The formulation phase transforms these examples into structured natural language specifications using the Gherkin language \cite{zhang2025think}. 
Gherkin is a domain-specific language created to support BDD \cite{zameni2023bdd}. It uses a plain-text, structured format that is easily readable by both technical and non-technical stakeholders. The core syntax for Gherkin is ‘Given-When-Then’. ‘Given’ sets up a precondition, ‘When’ describes the actions, and ‘Then’ specifies the expected outcomes \cite{rahman2015reusable}. Finally, the automation phase converts these specifications into executable test cases that verify the system behaves as expected \cite{zameni2023bdd}.
BDD offers several significant benefits to software development teams. It creates a universal language that all stakeholders can understand. It focuses development on delivering business value through user-centred features. It also provides living documentation that remains relevant throughout the project lifecycle. In addition, it encourages collaboration across disciplines and reduces misinterpretations of requirements.

\subsection {Challenges in BDD Implementation}
Despite its advantages, BDD faces several challenges that impact its effectiveness and adoption. Zameni et al. \cite{zameni2023bdd} note that BDD scenarios, while written in natural language, frequently lack sufficient precision for unambiguous interpretation. This creates difficulties when translating scenarios into executable tests.
Furthermore, Rahman et al.  \cite{rahman2015reusable} identify that BDD adoption requires considerable organisational change, as it demands collaboration across traditionally siloed departments. This cultural shift represents a significant barrier to implementation in many organisations \cite{irshad2021adapting}.
Specific challenges in BDD scenario creation include:
\begin{enumerate} [leftmargin=4ex]
\item Domain Knowledge Requirements: Creating an effective scenario requires a deep understanding of both the business domain and technical implementation details, a rare combination of skills \cite{karpurapu2024comprehensive}.

\item Quality and Consistency Issues: Karpurapu et al. \cite{karpurapu2024comprehensive} observed that creating effective BDD acceptance tests demands considerable practical experience, leading to inconsistent quality across different team members.
\item Time-Intensive Process: Manual scenario creation is labour-intensive, particularly for complex features with numerous edge cases and business rules \cite{zameni2023bdd}.
\item Maintenance Burden: As systems evolve, maintaining and updating existing scenarios requires significant effort, especially in large codebases \cite{chemnitz2023towards}.

\item Scenario Coverage Gaps: Manually created scenarios often focus on happy paths. These challenges create a bottleneck in the software development process, where the benefits of BDD are recognised, but practical implementation remains difficult to scale effectively.
\end{enumerate}
\subsection {Traditional Approaches to BDD Scenario Generation}

Prior to LLM-based approaches, several methodologies have been explored to address the challenges in BDD scenario creation. Zameni et al. \cite{zameni2023bdd} introduced BDD Transition Systems (BDDTS) as a formal model for BDD scenarios. Their approach translates Gherkin scenarios into formal models, composes these models for comprehensive test coverage, and then automatically generates and executes tests. While promising, this method still requires manual translation of scenarios into formal models, which demands technical expertise. Lafi et al. \cite{lafi2021automated} developed a system for automated test case generation that extracts data from use case diagram descriptions and generates corresponding control flow graphs and NLP tables to produce test paths. However, this approach is not specifically tailored for BDD scenario generation and requires substantial technical input. 

These traditional approaches, while valuable in specific contexts, highlight the need for more accessible, automated solutions that can reduce the expertise barrier while improving scenario quality and coverage.

\subsection {LLMs in Software Testing}
Recent research has begun exploring the potential of LLMs for various software testing tasks \cite{wang2024software,hou2024large}. Wang et al. \cite{wang2024software} found that program repair, system-level test input generation, and unit test case generation are the most common testing tasks for which LLMs have been utilised. For example, Moradi et al. \cite{dakhel2024effective} introduced MuTAP (Mutation Test case generation using Augmented Prompt), an approach that iteratively improves LLM-generated test cases by augmenting prompts and applying syntax/behavioural refinements, achieving superior bug detection compared to automated tools. In another study, Kathiresan \cite{kathiresan2024automated} proposed an AI-driven framework for automated test case generation using deep learning, reinforcement learning, and evolutionary algorithms to dynamically generate and prioritise test cases, improving test coverage, reducing manual effort, and enabling earlier defect detection in the software development cycle. Liu et al. \cite{liu2025can} introduced an LLM-based regression test generation technique that generates bug-revealing and bug-reproduction test cases for software commits in programs with human-readable inputs.

A close work to our research is the work done by Karpurapu et al. \cite{karpurapu2024comprehensive}. They evaluated multiple LLMs to generate BDD acceptance tests, demonstrating that GPT-3.5 and GPT-4 achieved superior syntactic correctness with few-shot prompting. Our work differs from Karpurapu et al. \cite{karpurapu2024comprehensive}: (1) Scale: 500 real-world industrial user stories with detailed requirements from four software products versus their $\approx$50 user stories from public sources. (2) Evaluation: multi-dimensional assessment (text similarity, semantic similarity, LLM-based, and human expert evaluation versus their syntactic validation alone. (3) Research Questions: systematic investigation of four dimensions (baseline effectiveness, three prompting techniques, input type variations, and model parameters) versus their focus on zero-shot versus few-shot prompting only.

\section{Methodology} \label{sec:method}
This section details the research methodology, including research questions, dataset curation, LLMs, prompt techniques, and evaluation methods.

\subsection{Research Questions}
Our research seeks to answer the following research questions.

\begin{center}
\label{sec:RQ1}
\begin{tcolorbox}[arc=1mm,width=1.0\columnwidth,
                  top=1mm,left=1mm,  right=1mm, bottom=1mm,
                  boxrule=.75pt]
{\textbf{RQ1}: How effective are LLMs in generating BDD scenarios?}
\end{tcolorbox}
\end{center}
\textbf{Rationale}: RQ1 assesses the effectiveness of LLMs in generating BDD scenarios. Given that current research in this area remains limited, there is insufficient evaluation of effectiveness across multiple quality dimensions. Establishing the baseline effectiveness of LLMs is important before exploring optimisation strategies. This question will provide evidence on whether LLMs can produce scenarios that meet standards by evaluating metrics, including automatic evaluation metrics (i.e., text similarity, semantic similarity, and LLM-based) and human evaluation.

\begin{center}
\label{sec:RQ2}
\begin{tcolorbox}[arc=1mm,width=1.0\columnwidth,
                  top=1mm,left=1mm,  right=1mm, bottom=1mm,
                  boxrule=.75pt]
{\textbf{RQ2}: What is the impact of different prompting techniques on the ability of LLMs to generate BDD scenarios?}
\end{tcolorbox}
\end{center}
\textbf{Rationale}: Prior studies (e.g., \cite{sun2024source,zeng2025evaluating}) have shown that LLM performance is highly dependent on the prompting strategy used. However, the identification of effective prompting techniques specifically for BDD scenario generation remains underexplored. This research investigates three prompting approaches, zero-shot, few-shot, and Chain-of-Though (CoT), to examine how each influences the quality and effectiveness of BDD scenarios generated by LLMs.

\begin{center}
\label{sec:RQ3}
\begin{tcolorbox}[arc=1mm,width=1.0\columnwidth,
                  top=1mm,left=1mm,  right=1mm, bottom=1mm,
                  boxrule=.75pt]
{\textbf{RQ3}: How do different input types influence the ability of LLMs to generate BDD scenarios?}
\end{tcolorbox}
\end{center}
\textbf{Rationale}: In real-world software development, the availability and quality of input information vary significantly. Teams may have detailed user stories, comprehensive requirement descriptions, or both, depending on their development maturity and documentation practices. Understanding how different input types affect the quality of scenarios has direct implications for generating BDD scenarios. This question addresses a critical gap in current research, which has not evaluated how input types influence LLMs to generate BDD scenarios. The findings will provide recommendations on information requirements for effective BDD scenario generation, helping practitioners prepare their input processes and set realistic expectations based on their available documentation.
\begin{center}
\label{sec:RQ4}
\begin{tcolorbox}[arc=1mm,width=1.0\columnwidth,
                  top=1mm,left=1mm,  right=1mm, bottom=1mm,
                  boxrule=.75pt]
{\textbf{RQ4}: How do different model settings influence the ability of LLMs to generate BDD scenarios?}
\end{tcolorbox}
\end{center}
\textbf{Rationale}: Model settings such as temperature and top\_p influence the stochastic behaviour of LLMs \cite{sun2024source} and may affect the quality and consistency of generated BDD scenarios. This research question examines the impact of different model settings on scenario generation and identifies configurations that balance high-quality output with reliable, reproducible results.

\begin{figure*}
    \centering
    \includegraphics[width= 1\linewidth]{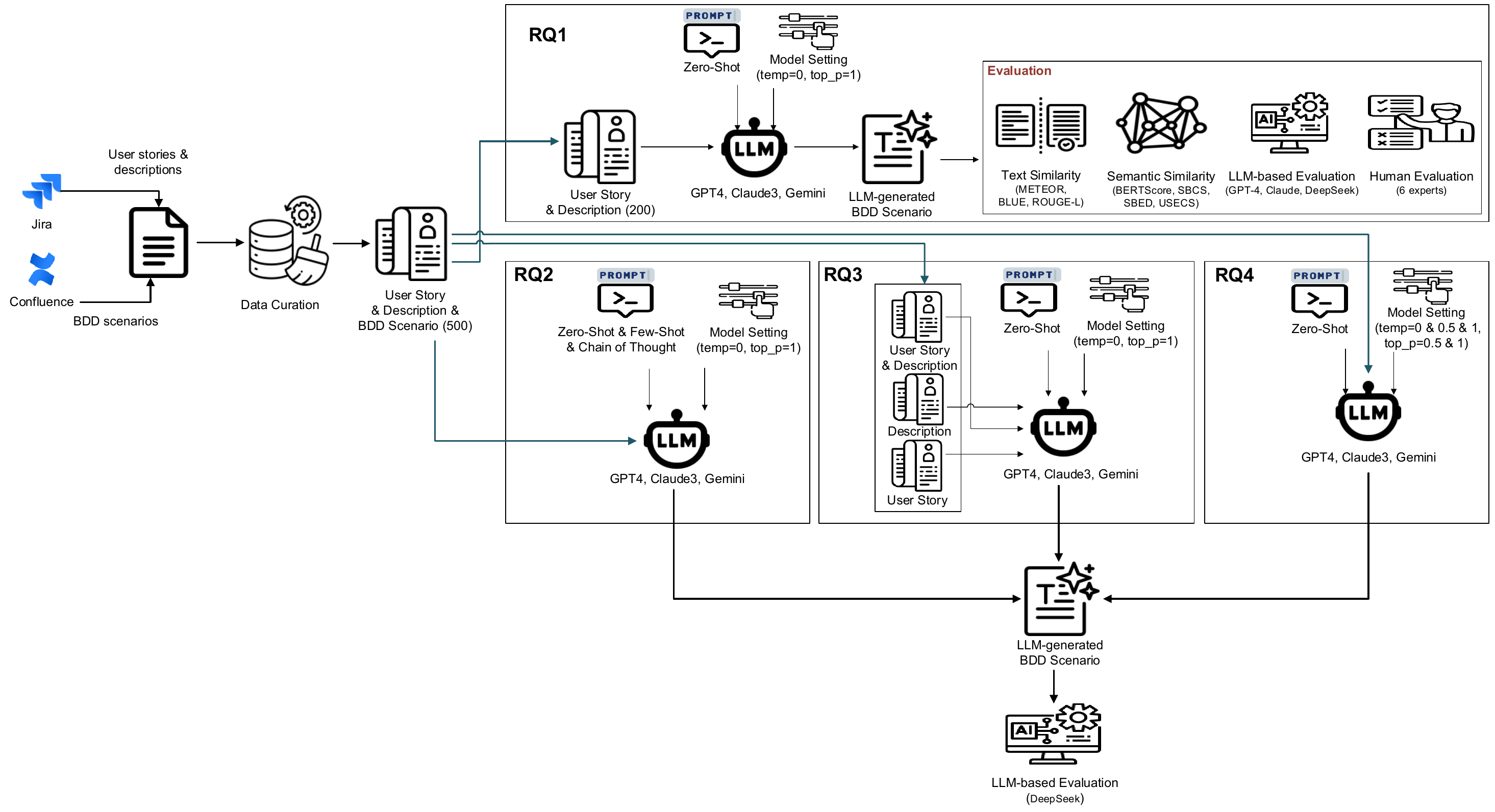}
   \caption{Overview of our methodology}
    \label{fig:BDD}
\end{figure*}

To answer these questions, we used three LLMs (GPT-4, Claude 3, Gemini) to generate BDD scenarios and evaluated them using four metrics (text similarity, semantic similarity, LLM-based evaluation, human evaluation). In the following, we first elaborate on the construction of a dataset and then explain experiments with three LLMs. Figure \ref{fig:BDD} presents an overview of our methodology in this study. The figure illustrates the main phases of the research process, including data collection, experimental design, model configuration, and evaluation procedures.

\subsection{Dataset Curation (First Contribution)} \label{sec:dataset}

As far as we are aware, there is no publicly available dataset of BDD scenarios that can be used to assess the effectiveness of LLMs in generating BDD scenarios. So, we had to construct a dataset of BDD scenarios. The dataset for this study was sourced from user stories and descriptions (requirements) in IntelligenceBank with formal company permission and appropriate data-sharing agreements in place. IntelligenceBank is a leading software company in Digital Asset Management and Marketing Operations Platform, operating in 55 countries. The company employs more than 100 employees with offices in Australia, the USA, Canada, and Europe. 

\subsubsection {Data Source and Initial Collection.}
User stories, descriptions, and their corresponding BDD scenarios were extracted from the company's project management tools. User stories and descriptions were collected from Jira\footnote{\href{https://www.atlassian.com/software/jira/}{https://www.atlassian.com/software/jira}}, while BDD scenarios were retrieved from Confluence\footnote{\href{https://www.atlassian.com/software/confluence/}{https://www.atlassian.com/software/confluence}} using Jira ID as the query parameter. The initial dataset was collected from four different software products with various team sizes and project durations, as shown in Table \ref{table:projectSummary}. For this research, we selected 500 user stories, their descriptions, and corresponding BDD scenarios from these software products. As shown in Table \ref{table:projectSummary},  we used a stratified random sampling strategy to sample use stories proportionally based on the software products' complexity, rather than selecting equal numbers of user stories from each product \cite{thompson2012sampling}. This approach ensured our sample effectively represented the diversity across these key dimensions while maintaining methodological integrity.

\begin{table*}[h]
 \caption{A summary of the four software products used to curate the dataset}
    \centering
    \begin{tabular}{|c|c|c|c|c|}
        \hline
\textbf{Software Products} & \textbf{Language} & \textbf{Team Size} & \textbf{Duration (Months)} & \textbf{Selected User Stories} \\ \hline
Digital Asset Management        & React/Elixir  & 12--15 & 18 & 150 out of 3500 \\ \hline
Brand Management                & PHP/Elixir    & 8--12  & 12 & 125 out of 2500\\ \hline
Marketing Operations Platform   & React/Elixir  & 15--18 & 24 & 150 out of 3000 \\ \hline
Marketing Compliance            & Python        & 6--8   & 8  & 75 out of 1000  \\ \hline
    \end{tabular}
  
    \label{table:projectSummary}
\end{table*}

Table \ref{table:dataset} presents an example illustrating the transformation of raw Jira data to the curated format. The user stories and descriptions were genuine business needs, stakeholder interaction and reflected company practices. Hence, we believe that the dataset is reliable, represents real software development artifacts, and has great potential for reuse by research and practice communities. We provided open access to our complete dataset and source code \cite{rathnayake2026llmbdd} to facilitate validation and further development of our research findings in subsequent studies.

\subsubsection {Data Curation Process.}
The raw data extracted from Jira and Confluence contained several quality issues that required systematic processing. We implemented the following systematic process to address dataset quality issues.
\begin{enumerate} [leftmargin=4ex]
\item {Metadata and Navigation Cleanup:} We removed Jira-specific elements that were not relevant to requirement content, such as system timestamps and user attribution tags (e.g., “Created by @username on 2023-05-15”), internal navigation links (e.g., “Related Issues”), status change notifications and workflow metadata, attachment references, and file links. 
\item {Comment and Discussion Removal:} We filtered out conversational elements to retain only the core requirement specification, such as comment threads between team members, review feedback and approval discussions, and status update annotations. 
\item {Format Standardization:} We established unified formatting rules across all descriptions, such as converted all bullet points to numbered lists (1., 2., 3., etc.), standardised indentation to a single level (removed nested structures) and unified terminology (e.g., standardised “user” vs “customer”, “system” vs “application”). 
\end{enumerate}

\begin{table}[htbp]
\centering
\caption{Data Transformation Example: Raw to Curated}
\small
\begin{tabular}{|p{0.95\columnwidth}|}
\hline
\textbf{Original Jira Export} \\
\hline
\textcolor{gray}{\textit{[Jira-2847] created by @firstname.lastname on 2025-04-12}} \\
\textbf{User Story:} As a content manager, I want to apply time-bound permissions \\
\textbf{Requirement Description:} \\
\hspace{0.5em}• Users should be able to set expiration dates \\
\hspace{1.5em}- Dates must be in future \\
\hspace{1.5em}- System validates date format \\
\hspace{0.5em}• Send notification before expiry \\
\textcolor{gray}{\textit{Related: Jira-2801, Jira-2756}} \\
\textbf{Comments (3):} \\
\hspace{0.5em}@firstname.lastname: Should we support timezone? \\
\hspace{0.5em}@firstname.lastname: Yes, UTC \\
\hline
\textbf{Curated Format} \\
\hline
\textbf{User Story:} As a content manager, I want to apply time-bound permissions to shared files so that external users can only access files during a specified timeframe \\
\textbf{Requirement Description:} \\
1. Users can set expiration dates for file permissions \\
2. Expiration dates must be in the future \\
3. System validates date format and timezone (UTC) \\
4. System sends notification 24 hours before permission expiry \\
\hline
\textbf{Confluence BDD Scenario} \\
\hline
\textbf{Scenario:} Set valid future expiration date for file permissions \\
\textbf{Given} I am logged in as a content manager \\
\hspace{0.5em}\textbf{And} I have a file named "Financial\_Report.pdf" in my library \\
\textbf{When} I navigate to the file sharing settings for "Financial\_Report.pdf" \\
\hspace{0.5em}\textbf{And} I share the file with external user "partner@external.com" \\
\hspace{0.5em}\textbf{And} I set the permission expiration date to "2025-12-31T23:59:59Z" \\
\hspace{0.5em}\textbf{And} I save the permission settings \\
\textbf{Then} the file should be shared with "partner@external.com" \\
\hspace{0.5em}\textbf{And} the permission expiration date should be "2025-12-31T23:59:59Z" \\
\hspace{0.5em}\textbf{And} the timezone should be stored as "UTC" \\
\hspace{0.5em}\textbf{And} a confirmation message "Permission successfully set with expiration date" should be displayed \\
\hspace{0.5em}\textbf{And} a notification should be scheduled for "2025-12-30T23:59:59Z" \\
\hline
\end{tabular}
\label{table:dataset}
\end{table}

\subsection{LLMs} \label{sec:LLMs}

In this research, we chose three LLMs as representative models, considering various sizes and model families. These LLMs have been extensively used to solve software engineering tasks such as code summarisation, code completion, and test case generation \cite{hou2024large,fan2023large}.

\textbf{GPT-4} (Generative Pre-trained Transformer) \cite{openai2025api} represents a large language model developed by OpenAI. Trained on vast collections of texts and code \cite{openai2025gpt4}, it demonstrates capabilities in understanding and creating natural language and code while resolving difficult problems more accurately \cite{sun2024source}. Our study used gpt-4o version.

\textbf{Gemini} \cite{team2023gemini} is a series of multimodal systems that leverage Transformer decoder frameworks and are trained on data across different modalities and language types. The Gemini family is structured into three different sizes: Pro, Flash, and Flash-Lite. We used the gemini-1.5-flash model, a well-established large language model in evaluation scenarios, with the capacity to manage contexts containing 10M tokens \cite{xue2024automated}, and we used Google APIs for this study.

\textbf{Claude 3} \cite{anthropic2025claude3haiku} stands as the latest evolution of the Claude intelligence architecture, engineered by Anthropic. The family includes three models: Haiku, Sonnet, and Opus. Our study used claude-3-opus-20240229. For this research, the Claude Opus model was chosen based on its exceptional accuracy, quick responses to sophisticated inquiries, and its position as the most advanced intelligence model \cite{anthropic2025claude3modelcard}.

\subsection {Prompting Techniques} \label{sec:Prompts}
In our research, we employed three commonly used prompting techniques \cite{hou2024large}. 

\begin{figure}
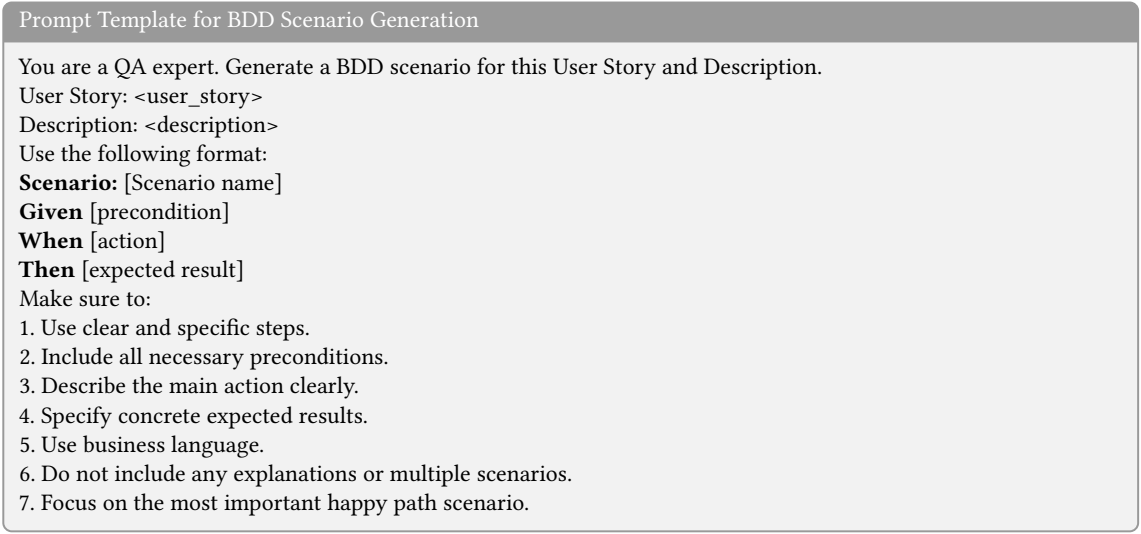

\begin{tcolorbox}[arc=1mm,width=1.0\columnwidth,
                  top=1mm,left=1mm,  right=1mm, bottom=1mm,
                  boxrule=.75pt, colback=gray!10, colframe=gray!80, title=Prompt Template for BDD Scenario Generation]

You are a QA expert. Generate a BDD scenario for this User Story and Description.

User Story: <user\_story>

Description: <description>

Use the following format:

\textbf {Scenario:} [Scenario name]

\textbf {Given} [precondition]

\textbf {When} [action]

\textbf {Then} [expected result]
        
Make sure to:

1. Use clear and specific steps.

2. Include all necessary preconditions.

3. Describe the main action clearly.

4. Specify concrete expected results.

5. Use business language.

6. Do not include any explanations or multiple scenarios. 

7. Focus on the most important happy path scenario.

  \end{tcolorbox}
  \caption{Zero-Shot Prompt}
  \label{fig:ZeroShotPrompt}
\end{figure}

\textbf {Zero-Shot.} Through minimal instructional guidance, zero-shot prompting adjusts LLMs for downstream task execution \cite{sun2024source}. In our scenario generation process, the LLMs are given clear directions along with the user story and description (requirements), as shown in Figure \ref{fig:ZeroShotPrompt}. The expected output is natural language scenarios structured in the Gherkin format. 

\textbf {Few-Shot.} Few-shot prompting offers clear directives combined with sample cases for adapting LLMs to downstream tasks \cite{geng2024large,gao2023makes}. The provided examples serve as contextual conditioning for later scenarios that require LLM-generated responses \cite{sun2024source}. The few-shot prompt template used in this study is shown in Figure \ref{fig:FewShotPrompt}.

\begin{figure}
\begin{tcolorbox}[arc=1mm,width=1.0\columnwidth,
                  top=1mm,left=1mm,  right=1mm, bottom=1mm,
                  boxrule=.75pt, colback=gray!10, colframe=gray!80, title=Prompt Template for BDD Scenario Generation]

You are a QA expert. Generate a BDD scenario for the given user story and description using the examples below as guidance.
 
\textbf {Example 1:}

\textbf {\textit {User Story:}} As a marketing manager, I want to upload brand assets to the system so that my team can access approved materials.

\textbf {\textit{Description:}} The system should allow users with appropriate permissions to upload image files (PNG, JPG, JPEG) up to 10MB in size. The uploaded files should be automatically categorised and made available to team members.

\textbf {\textit {BDD Scenario:}} Marketing manager uploads brand logo successfully

Given I am logged in as a marketing manager with upload permissions

And I am on the asset upload page

When I select a PNG file that is 5MB in size

And I click the "Upload" button

Then the file should be uploaded successfully

And I should see a confirmation message "File uploaded successfully"

And the file should appear in my brand assets library

\hfill \break
\textbf {Example 2:}

\textbf {\textit {User Story:}} As a content creator, I want to search for existing assets by keyword so that I can quickly find relevant materials for my projects.

\textbf {\textit {Description:}} Users should be able to enter keywords in a search box and receive filtered results showing matching assets. The search should work across file names, tags, and descriptions with results displayed in a grid format.

\textbf {\textit {BDD Scenario:}} Content creator searches for logo assets

Given I am logged in as a content creator

And there are assets tagged with "logo" in the system

When I enter "logo" in the search box

And I click the search button

Then I should see a list of assets matching "logo"

And the results should be displayed in a grid format

And each result should show the asset thumbnail and name

\textbf {Your Task:} 
Now generate a BDD scenario for this user story and description:

User Story: <user\_story>

Description: <description>

\textbf {Output Format:}

\textbf {Scenario:} [Scenario name]

\textbf {Given} [precondition]

\textbf {When} [action]

\textbf {Then} [expected result]
        
Make sure to:

1. Use clear and specific steps.

2. Include all necessary preconditions.

3. Describe the main action clearly.

4. Specify concrete expected results.

5. Use business language.

6. Do not include any explanations or multiple scenarios. 

7. Focus on the most important happy path scenario.

  \end{tcolorbox}
  \caption{Few-Shot Prompt}
  \label{fig:FewShotPrompt}
\end{figure}

\textbf {Chain-of-Thought (CoT).} LLMs are trained for downstream tasks via chain-of-thought prompting, which encourages step-by-step reasoning \cite{wei2022chain}. These steps equip LLMs with advanced reasoning abilities. Following Wang et al. \cite{wang2023element}, we applied chain-of-thought prompting to the BDD scenario generation task in three steps: identify core elements, choose a primary scenario, and structure the scenario. The CoT prompting technique used in our research is shown in Figure \ref{fig:CoTprompt}.

\begin{figure}
  \begin{tcolorbox}[arc=1mm,width=1.0\columnwidth,
                  top=1mm,left=1mm,  right=1mm, bottom=1mm,
                  boxrule=.75pt, colback=gray!10, colframe=gray!80, title=Prompt Template for BDD Scenario Generation]
You are a QA expert specialising in Behaviour-Driven Development. Generate a comprehensive BDD scenario for the User Story and Description.

User Story: <user\_story>

Description: <description>

\textbf {Analysis Steps:}

\textbf {Step 1: Identify Core Elements.}

\begin{itemize}
\item Main user goal and business value
\item Key user actions and system interactions
\item Success criteria and potential failure points
\end{itemize}

\textbf {Step 2: Choose Primary Scenario.}
\begin{itemize}
\item Focus on the main success path that delivers core business value
\item Consider the most important user interaction
\end{itemize}

\textbf {Step 3: Scenario Structure.}
\begin{itemize}
\item Given: Specific system state and user context
\item When: Clear, testable user actions
\item Then: Measurable, observable outcomes
\end{itemize}

\textbf {Output Format:}

\textbf {Scenario:} [Scenario name]

\textbf {Given} [precondition]

\textbf {When} [action]

\textbf {Then} [expected result]
        
Make sure to:

1. Use clear and specific steps.

2. Include all necessary preconditions.

3. Describe the main action clearly.

4. Specify concrete expected results.

5. Use business language.

6. Do not include any explanations or multiple scenarios. 

7. Focus on the most important happy path scenario.

  \end{tcolorbox}
  \caption{Chain-of-thought Prompt}
  \label{fig:CoTprompt}
\end{figure}

\subsection {Evaluation Metrics} \label{sec:EvaluationMetrics}
This section discusses the different evaluation metrics used in this research to evaluate the LLMs, including automated and human evaluations. 

\subsubsection{Automated Evaluation Metrics.}
The automated metrics for LLM-generated scenarios can be divided into the following three categories.

\textbf{Text Similarity Metrics}: The metrics in this category evaluate the quality of the generated scenario by quantifying text similarity between the generated scenario and its reference counterpart. Existing text evaluation research frequently uses this category of evaluation methods \cite{geng2024large,gao2023makes}. Among existing text similarity metrics, we employed Metric for Evaluation of Translation with Explicit Ordering (METEOR), Bilingual Evaluation Understudy (BLUE), and Recall-Oriented Understudy for Gisting Evaluation - Longest Common Subsequence (ROUGE-L), each analysing n-gram distribution similarities between generated and reference scenarios \cite{sun2024source}. 
The scoring scale for METEOR, BLUE, and ROUGE-L ranges from 0 to 1 inclusively \cite{sun2024source}. As the score increases, it reflects a close match between the generated scenario and the reference scenario. 

\textbf{Semantic Similarity Metrics}: The metrics in this category assess the quality of the generated scenario by measuring the semantic relatedness between the generated scenario and its reference scenario. According to previous studies \cite{haque2022semantic}, semantic similarity metrics successfully address the limitations of word overlap metrics, particularly their failure to recognise varying word significance and synonymous relationships. In our research, four semantic similarity metrics are used: BERTScore \cite{zhang2019bertscore}, SentenceBert with Cosine Similarity (SBCS), SentenceBert with Euclidean Distance (SBED), and Universal Sentence Encoder \cite{cer2018universal} with Cosine Similarity (USECS). 

Utilising a modified version of Bidirectional Encoder Representations from Transformers (BERT) \cite{devlin2019bert}, BERTScore \cite{zhang2019bertscore} embeds individual tokens within scenarios and determines the pairwise inner products between reference scenario tokens and those produced by LLMs \cite{sun2024source}. Next, tokens from both the reference and LLM-generated scenarios undergo matching to quantify precision, recall, and F1 values. The F1 value resulting from BERTScore analysis is presented in our research outcomes. Pre-trained sentence encoders (SentenceBERT \cite{reimers2019sentence} or Universal Sentence Encoder \cite{cer2018universal}) are utilised in the other methods to transform scenarios into vector formats; subsequently, mathematical proximity is determined using cosine similarity or Euclidean distance metrics. The value range for SBCS, SBED, and USECS spans from -1 to 1. Greater SBCS and USECS measurements reflect enhanced similarity, while diminished SBED figures indicate closer correspondence \cite{sun2024source}.

\textbf{LLM-based Evaluation}: Motivated by recent work \cite{wang2023chatgpt,vykopal2024disinformation,liu2023g}, we investigated the feasibility of employing LLMs as evaluators to assess the quality of the BDD scenarios. Specifically, we employed two of the LLMs used for BDD scenario generation (see Section~\ref{sec:LLMs}) as evaluators, and included DeepSeek as a third LLM-based evaluator. DeepSeek \cite{liu2024deepseek} is a strong Mixture-of-Experts (MoE) language model with 671B total parameters. We selected DeepSeek-R1 \cite{guo2025deepseek}, which is trained to give thought processes ahead of task answers, advancing problem-solving performance. Each LLM evaluator receives user stories, descriptions, and LLM-generated BDD scenarios. The LLM evaluators assess each scenario on a 5-point rating scale, with 5 representing the highest quality. Figure \ref{fig:llm_evaluation_prompt} illustrates the prompt used for LLM evaluation.

\begin{figure}
  \begin{tcolorbox}[arc=1mm,width=1.0\columnwidth,
                  top=1mm,left=1mm,  right=1mm, bottom=1mm,
                  boxrule=.75pt, colback=gray!10, colframe=gray!80, title=Prompt Template for Evaluating Generated BDD Scenario]
You are an expert software testing evaluator specialising in Behaviour-Driven Development (BDD) scenarios. Your task is to evaluate the BDD scenario against the provided requirements (both user story and description) and rate it on a scale from 1 to 5, where a higher score indicates better quality.

\textbf {Evaluation Context:} User Story, Description, and BDD Scenario.

\textbf {Evaluation Instructions:} When evaluating the BDD scenario, consider ALL provided requirements. A good BDD scenario should have:

\begin{itemize}
\item Clear, specific, and testable steps
\item Comprehensive coverage of ALL provided requirements
\item Excellent clarity and precision in test assertions
\item Proper Gherkin syntax and BDD structure
\end{itemize}

\textbf {Response Format:} Please respond in the following format.
\begin{itemize}
\item Rating: <1, 2, 3, 4, or 5>
\item Reasoning: Detailed explanation of your evaluation, including specific strengths and weaknesses.
\end{itemize}

\end{tcolorbox}
\caption{Prompt used in LLM-based evaluation}
\label{fig:llm_evaluation_prompt}
\end{figure}
\subsubsection{Human Evaluation.} \label{sec:human_evaluation}

In this research, we also conducted a human evaluation to assess the quality of the scenarios generated by LLMs. The results of this evaluation can inform the design of experiments to identify automated metrics that strongly and positively correlate with human judgments (RQ1). Such metrics could ultimately serve as reliable substitutes for human evaluation in large-scale datasets. The human evaluation process was conducted by six senior QA professionals (2 males and 4 females), each with more than a decade of software experience in BDD and strong English-language proficiency. The following steps describe the process.

\textbf {Step 1: Sample Selection.} As described in Section \ref{sec:dataset}, our dataset comprises 500 user stories with their corresponding descriptions. Using three LLMs (see Section \ref{sec:LLMs}), we generated BDD scenarios for each user story and its description, resulting in a total of 1,500 scenarios. To balance comprehensive evaluation with a manageable workload for human evaluators, we sampled a subset of these 1,500 scenarios. Specifically, from the 1,500 generated scenarios, we selected 600 for human evaluation, 200 scenarios per LLM. These 200 scenarios were derived from 200 user story–description pairs across the four software projects, with 50 samples per project. Note that the 1500 scenarios selected were those produced during the third experimental run. To ensure unbiased and reliable assessment, the evaluation workload was distributed across the six evaluators. The 600 scenarios were evaluated independently by two evaluators.
\begin{itemize} [leftmargin=4ex]
\item 200 Claude 3's generated BDD scenarios: Evaluators 1 and 2
\item 200 GPT-4's generated BDD scenarios: Evaluators 3 and 4
\item 200 Gemini's generated BDD scenarios: Evaluators 5 and 6
\end{itemize}
The evaluators received evaluation materials consisting of use stories, descriptions, and the corresponding LLM-generated BDD scenarios.

\textbf {Step 2: Assessment.}
Inspired by Shi et al. \cite{shi2022evaluation}, we instructed evaluators to rate each of the 200 scenarios assigned to them on a 5-point scale, where 5 indicates the highest quality (the scenario perfectly meets the requirements), and 1 indicates the lowest quality.

\textbf {Step 3: Resolving Disagreements.} The next step involved comparing the ratings from the two evaluators assigned to each scenario to identify discrepancies. Our analysis revealed that the evaluators had 17 disagreements in the scenarios generated by Claude 3, 23 in those generated by GPT-4, and 25 in those generated by Gemini. Next, we asked a designated moderator, chosen from the evaluators, to resolve disagreements and reach agreement on the final ratings. This consensus process ensured that the final human evaluation scores reflect collective expert judgment.

\section{Findings}\label{sec:findings}

This section presents the findings from our experiments using LLMs to generate BDD scenarios under different conditions to answer RQ1, RQ2, RQ3, and RQ4. All experimental setups have been shown in Figure \ref{fig:BDD}.

\subsection{RQ1: How effective are LLMs in generating BDD scenarios?}

\subsubsection {Experimental Setup.}
As shown in Figure \ref{fig:BDD}, in RQ1, all three LLMs were provided with \textit{user stories} and \textit{descriptions} as inputs and used zero-shot prompting to generate BDD scenarios. We also set the temperature to 0 and top\_p to 1 for all LLMs to produce deterministic and consistent outputs. We used a comprehensive evaluation strategy that combined automated metrics (text similarity, semantic similarity, and LLM-based evaluation) and human evaluation (See Section \ref{sec:EvaluationMetrics}). Because RQ1 also involved human evaluation (as described in Section \ref{sec:human_evaluation}), the sample size was limited to 200 to reduce the workload of human evaluators. To compute automated metrics and improve reliability, each experiment (i.e., BDD scenario generation) was repeated three times. The final score of the automated metrics for each scenario was calculated as the average across the three runs. For human evaluation, the BDD scenarios generated in the third run were selected, as described in Section \ref{sec:human_evaluation}.

\subsubsection {Experimental Results.}\label{sec:RQ1findings}
Table \ref{tab:evaluation-comparison} presents the results of applying the three LLMs to generate BDD scenarios from \textit{user stories} and \textit{descriptions}, evaluated using four metrics. The results indicate that while all three models can produce BDD scenarios, the quality of the generated scenarios varies. 

\begin{table*}[htbp]
\caption{Assessing the effectiveness of LLMs in BDD scenarios generation with different evaluation metrics. ZS: Zero-Shot. The score of the best-performing LLM for each metric is highlighted in \textbf{bold}.}
\centering
\setlength{\tabcolsep}{3pt} 
\small 
\begin{tabular}{ll|ccc|cccc|ccc|c}
\hline
\begin{tabular}[c]{@{}c@{}}\textbf{LLMs}\end{tabular} & \begin{tabular}[c]{@{}c@{}}\textbf{Prompt}\end{tabular} & \multicolumn{3}{c|}{\textbf{Text Similarity}} & \multicolumn{4}{c|}{\textbf{Semantic Similarity}} & \multicolumn{3}{c|}{\textbf{LLM Evaluation}}& \begin{tabular}[c]{@{}c@{}}\textbf{Human}\end{tabular}\\ \hline
\cline{3-5} \cline{6-9} \cline{10-12}
 &  & BLEU & METEOR & ROUGE-L & BERTScore & SBCS & SBED & USECS & DeepSeek & GPT-4 & Claude \\ \hline
Gemini & ZS & 7.91 & 32.15 & 34.74  & 89.46  & 51.06  & 39.68  & 45.95  & 3.72 & 4.19 & 3.48 & 3.39\\ \hline
Claude 3 & ZS & 11.02  & 32.12  & 38.36 & 90.75 & 53.62 & 41.85 & 50.53 & \textbf{4.04} & \textbf{4.47} & \textbf{3.79} & \textbf{4.06}\\ \hline
GPT-4 & ZS & \textbf{12.37} & \textbf{36.23} & \textbf{40.63} & \textbf{91.16} & \textbf{53.96} & \textbf{42.16} & \textbf{51.95} & 3.55 & 4.38 & 3.78 & 3.79\\ \hline

\end{tabular}
\label{tab:evaluation-comparison}
\end{table*}

\textbf{Automated Evaluation Results.} We used three types of automated metrics to assess the effectiveness of LLMs.

\underline{\textit{Text Similarity Results}}. Table \ref{tab:evaluation-comparison} shows that GPT-4 achieved the highest score in the three text similarity metrics, BLEU (12.37), METEOR (36.23), ROUGE-L (40.63), closely followed by Claude 3. In all metrics, Gemini was the least performing model. Overall, all LLMs achieved relatively low BLEU scores, whereas their METEOR and ROUGE-L scores were notably higher.

\underline{\textit{Semantic Similarity Results.}} The LLMs showed almost similar behaviour in the semantic similarity metrics as in the text similarity metrics. GPT-4 achieved the highest scores across all text similarity metrics, followed by Claude 3. Although all three LLMs achieved a high score in the BERTScore metric (>89\%), indicating strong capability for capturing BDD scenario meaning and context, their scores in other metrics, SBCS, SBED, and USECS, were comparatively lower.

\underline{\textit{LLM-based Evaluation Results.}} The LLM-based evaluation results in Table \ref{tab:evaluation-comparison} show a notable difference compared to the text and semantic similarity metrics. The DeepSeek evaluator ranked the scenarios generated by Claude 3 (4.04) higher than those generated by Gemini (3.72) and GPT-4 (3.55). GPT-4 as an evaluator was more optimistic, consistently assigning higher scores to the scenarios generated by Claude 3 (4.47), GPT-4 (4.38), and Gemini (4.19). The performance of Claude 3, as an evaluator, is almost identical to that of the other two LLM-based evaluators (DeepSeek and GPT-4), while scoring the scenarios generated by Gemini lower. Notably, all three LLM-based evaluators consistently ranked Claude 3's generated scenarios as the highest quality.

\textbf {Human Evaluation Results.}\label{sec:humanevalution} Table \ref{tab:evaluation-comparison} shows human evaluation results. The human evaluators ranked the scenarios generated by Claude 3 (4.06) higher than those generated by GPT-4 (3.79) and Gemini (3.39). These scores indicate that the generated scenarios were generally rated between \textit{good} and \textit{excellent} on the 5-point scale.

\begin{center}
\label{sec:RQ1}
\begin{tcolorbox}[arc=1mm,width=1.0\columnwidth,
                  top=1mm,left=1mm,  right=1mm, bottom=1mm,
                  boxrule=.75pt]
\ding{43} \textbf{Key Finding} $\blacktriangleright$ Our findings indicate that, although GPT-4 achieves higher performance in generating BDD scenarios according to text similarity and semantic similarity metrics, Claude 3 performs better from the perspectives of both LLM-based evaluation and human evaluation. $\blacktriangleleft$
\end{tcolorbox}
\end{center}

\subsubsection{Correlation between Automated Evaluation and Human Evaluation.} 
Human evaluation, described in Section \ref{sec:RQ1findings}, can be time-consuming, particularly as the dataset size increases. Inspired by \cite{wang2023chatgpt,liu2023g,sun2024source}, we were interested in understanding which of the automated evaluation metrics are positively and strongly correlated with human evaluation. To calculate the correlation between automated metrics (text similarity, semantic similarity, and LLM-based evaluation) and human evaluation, we used Spearman's correlation coefficient (ρ). Spearman's ρ ranges from -1 to 1. It shows correlations between ordered discrete or continuous data sequences, with larger absolute values indicating stronger relationships \cite{conover1999practical}. Values where $-1 \leq \rho < 0$ signify negative correlation, $\rho = 0$ indicates no correlation, and $0 < \rho \leq 1$ indicate positive correlation \cite{dancey2007statistics}. 

Table \ref{tab:correlation_results} shows the results of Spearman's correlation coefficients. We observe that text similarity metrics (BLEU, METEOR, ROUGE-L) and most semantic similarity metrics (SBCS, SBED, USECS) show weak correlations with human evaluation (ρ was less than 0.30 in all LLMs).  BERTScore achieves a moderate positive correlation with human evaluation (ρ = 0.40 to 0.48 across all LLMs). LLM-based evaluators consistently show stronger correlations with human evaluation than text similarity and semantic similarity metrics. Among GPT-4, Claude, and DeepSeek as LLM evaluators, the DeepSeek evaluator achieved the highest Spearman's correlations: ρ = 0.62 (GPT-4), ρ = 0.72 (Claude), and ρ = 0.53 (Gemini). The consistent performance of the DeepSeek evaluator ensures reliable evaluation across our experimental conditions, while its independence from the quality of reference scenarios makes it practical for large-scale evaluation. Hence, in RQ2, RQ3 and RQ4, we employ DeepSeek as a substitute for human evaluators to assess the quality of BDD scenarios generated by LLMs.

\begin{table*}[t]
\centering
\small
\caption{Spearman's correlation between automatic metrics (text similarity, semantic similarity, LLM-based evaluation) and human evaluation. For each LLM, the metric with the highest correlation to human evaluation is highlighted in \textbf{bold}.}
\label{tab:correlation_results}
\begin{tabular}{@{}lcccccccccc@{}}
\toprule
\multirow{2}{*}{\textbf{LLMs}} & \multicolumn{3}{c}{\textbf{Text Similarity}} & \multicolumn{4}{c}{\textbf{Semantic Similarity}} & \multicolumn{3}{c}{\textbf{LLM Evaluation}} \\
\cmidrule(lr){2-4} \cmidrule(lr){5-8} \cmidrule(lr){9-11}
& \textbf{BLEU} & \textbf{METEOR} & \textbf{ROUGE-L} & \textbf{BERTScore} & \textbf{SBCS} & \textbf{SBED} & \textbf{USECS} & \textbf{GPT-4} & \textbf{Claude 3} & \textbf{DeepSeek} \\
\midrule
GPT-4   & 0.08 & 0.16 & 0.30 & 0.48 & 0.15 & 0.18 & 0.06 & 0.57 & 0.61 & \textbf{0.62} \\
Claude 3  & 0.08 & 0.07 & 0.13 & 0.41 & 0.05 & 0.07 & 0.05 & 0.56 & 0.60 & \textbf{0.72} \\
Gemini  & 0.05 & 0.11 & 0.10 & 0.40 & 0.03 & 0.04 & 0.04 & 0.47 & 0.43 & \textbf{0.53} \\
\bottomrule
\end{tabular}
\end{table*}

\begin{center}
\begin{tcolorbox}[arc=1mm,width=1.0\columnwidth,
                  top=1mm,left=1mm,  right=1mm, bottom=1mm,
                  boxrule=.75pt]
\ding{43} \textbf{Key Finding} $\blacktriangleright$ Our experiments show that LLM-based evaluators correlate more strongly with human evaluations than text similarity and semantic similarity metrics. Among them, the DeepSeek evaluator exhibits the strongest positive correlation with human judgments. We also observe that most text similarity and semantic similarity metrics show weak correlations with human evaluation, with the exception of BERTScore.$\blacktriangleleft$
\end{tcolorbox}
\end{center}

\subsection{RQ2: What is the impact of different prompting techniques on the ability of LLMs to generate BDD scenarios?}

\subsubsection {Experimental Setup.}
For RQ2, we investigated how different prompting techniques influence the quality of BDD scenarios generated by LLMs. We evaluated three prompting techniques: zero-shot, few-shot, and Chain-of-Thought (CoT), as described in Section \ref{sec:Prompts}. As shown in Figure \ref{fig:BDD}, all three LLMs (see Section \ref{sec:LLMs}) were provided both \textit{user stories} and \textit{descriptions} as input to generate BDD scenarios using each of the three prompting techniques. Based on the findings from RQ1, we used DeepSeek as our LLM evaluator to assess the quality of the generated scenarios. DeepSeek evaluated all 500 BDD scenarios (see Section \ref{sec:dataset}) generated by each LLM under each prompting condition, using a 5-point rating scale where higher scores indicate better quality. To ensure reliability in LLM outputs, we conducted three independent evaluation runs for each configuration. The final evaluation score for each scenario was calculated as the average across the three runs. For this experiment, all LLMs were configured with temperature=0 and top\_p=1 to maintain consistency across prompting techniques.

\subsubsection {Experimental Results.}

\begin{table}[t]
\centering
\caption{The impact of different prompt techniques on LLM performance in generating BDD scenarios. The score of the best-performing LLM for each prompt technique is highlighted in \textbf{bold}. The score of the best-performing LLM across all prompt techniques is highlighted in \textbf{bold} and \underline{underlined}.} 
\label{tab:prompt techniques}
\begin{tabular}{lccc}
\toprule
\textbf{Model} & \textbf{Zero-Shot} & \textbf{Few-Shot} & \textbf{CoT} \\
\midrule
Claude 3 & 4.18  & 4.00 & 4.22 \\
GPT-4 & \textbf{\underline{4.63}} &  3.90 & \textbf{4.39} \\
Gemini &  4.17 & \textbf{4.34} & 4.14 \\
\bottomrule
\end{tabular}
\end{table}

Table \ref{tab:prompt techniques} presents the results of three LLMs across different prompting techniques, revealing distinct patterns in how each model responds. GPT-4 achieved its highest performance with zero-shot prompting (4.63), outperforming both few-shot (3.90) and Chain-of-Thought (4.39) techniques. This suggests that GPT-4 has strong capabilities for understanding requirements and generating high-quality BDD scenarios without requiring additional guidance through examples or explicit reasoning steps. 

Claude 3 demonstrated relatively consistent performance across all three prompting techniques, with chain-of-thought achieving the highest score (4.22), followed closely by zero-shot (4.18) and few-shot (4.00). The marginal improvement with CoT prompting suggests that Claude 3 benefits from explicit guidance on reasoning steps when generating BDD scenarios. The smaller performance variation across prompting techniques indicates that Claude 3 maintains stable quality regardless of the prompting technique employed. Gemini showed a different pattern, achieving its best performance with few-shot prompting (4.34), which outperformed both zero-shot and chain-of-thought. Its improvement from zero-shot to few-shot indicates that providing exemplar scenarios helps Gemini better understand the requirements and generate more appropriate BDD scenarios.

\begin{center}
\begin{tcolorbox}[arc=1mm,width=1.0\columnwidth,
                  top=1mm,left=1mm,  right=1mm, bottom=1mm,
                  boxrule=.75pt]
\ding{43} \textbf{Key Finding} $\blacktriangleright$ Our study shows that the impact of prompting techniques on the quality of BDD scenario generation varies significantly across LLMs. GPT-4 performs best with zero-shot prompting, Claude 3 shows marginal preference for chain-of-thought prompting, while Gemini achieves optimal results with few-shot prompting. Overall, the highest-quality BDD scenarios are produced by GPT-4 in a zero-shot prompting. $\blacktriangleleft$
\end{tcolorbox}
\end{center}

\subsection{RQ3: How do different input types influence the ability of LLMs to generate BDD scenarios?}
\subsubsection {Experimental Setup.}
For RQ3, we were interested in investigating how the three LLMs perform when provided with different information. Hence, we provided each LLM (1) user story only, (2) requirements description only, and (3) both the user story and description as input to generate BDD scenarios. DeepSeek evaluated all 500 BDD scenarios (see Section \ref{sec:dataset}) generated by each LLM with three input types using zero-shot prompting with a 5-point rating scale, where higher scores indicate better quality. We conducted three independent evaluation runs to ensure the reliability of LLM outputs and calculated the average presented in the Table \ref{tab:input_types}. As shown in Figure \ref{fig:BDD}, all LLMs were configured with temperature=0 and top\_p=1 to maintain consistency.

\begin{table}[t]
\centering
\caption{The impact of different inputs on LLM performance in generating BDD scenarios. The score of the best-performing LLM for each input is highlighted in \textbf{bold}. The score of the best-performing LLM across all inputs is highlighted in \textbf{bold} and \underline{underlined}.}
\label{tab:input_types}
\begin{tabular}{lccc}
\toprule
\textbf{Model} & \textbf{User Story + Description} & \textbf{Description Only} & \textbf{User Story Only} \\
\midrule
Claude 3 & 4.18 & 3.99 & 3.16 \\
GPT-4 & \textbf{\underline{4.63}} & \textbf{4.35} & 3.33 \\
Gemini & 4.17 & 4.07 & \textbf{3.37} \\
\bottomrule
\end{tabular}
\end{table}

\subsubsection {Experimental Results.}

Table \ref{tab:input_types} presents the performance of three LLMs when generating BDD scenarios using different input types. The results reveal significant variations in the quality of the scenario based on the type of input information provided.

\textbf {User Story + Description:} All three LLMs achieved their highest performance when provided with both user stories and descriptions. GPT-4 demonstrated the strongest performance (4.63), followed closely by Claude 3 (4.18) and Gemini (4.17). This indicates that combining user stories and descriptions enables LLMs to generate comprehensive, high-quality BDD scenarios that capture business and technical requirements.

\textbf {Description Only:} When provided with descriptions alone, all three LLMs maintained strong performance, with scores quite close to those achieved with the user story and description. GPT-4 scored 4.35 (a decrease of only 6.0\% from the combined input), Claude 3 achieved 3.99 (a decrease of 4.5\%), and Gemini scored 4.07 (a decrease of 2.4\%). These minimal performance degradations suggest that detailed descriptions contain sufficient technical specifications and contextual information to enable effective BDD scenario generation. 

\textbf {User Story Only:} All three LLMs showed performance degradation when provided with user stories alone. GPT-4's score dropped to 3.33 (a 28.1\% decrease from combined input), Claude 3 fell to 3.16 (a 24.4\% decrease), and Gemini declined to 3.37 (a 19.2\% decrease). These decreases indicate that user stories, despite providing valuable business context and user-centric perspectives, lack the technical depth and detailed specifications needed to generate comprehensive BDD scenarios. User stories typically follow the \textit{“As a [role], I want [feature], so that [benefit]”} format, which captures the high-level intent but misses specific acceptance criteria, edge cases, and detailed behaviour specifications that are essential for effective BDD scenarios.

\begin{center}
\label{sec:RQ2}
\begin{tcolorbox}[arc=1mm,width=1.0\columnwidth,
                  top=1mm,left=1mm,  right=1mm, bottom=1mm,
                  boxrule=.75pt]
\ding{43} \textbf{Key Finding} $\blacktriangleright$ We find that input quality fundamentally determines an LLM's effectiveness in BDD scenario generation. While LLMs perform best when provided with both user stories and requirement descriptions, detailed requirement descriptions, whether provided alone or combined with user stories, enable LLMs to generate quality BDD scenarios. In contrast, relying solely on user stories significantly degrades the quality of BDD scenarios. 
$\blacktriangleleft$
\end{tcolorbox}
\end{center}

\subsection {RQ4: How do different model settings influence the ability of LLMs to generate BDD scenarios?}
\subsubsection {Experimental Setup.}
Three key parameters (top\_k, top\_p, and temperature) control the degree of randomness in LLM-generated outputs \cite{sun2024source}. Since GPT-4 does not support the top\_k setting, we conducted experiments only with top\_p and temperature.

\textbf {Top\_p}: During each token generation cycle, LLMs rank tokens in descending order of probability and retain only those tokens whose cumulative probability reaches (but does not exceed) the top\_p threshold. For instance, setting top\_p=0.1 restricts consideration to tokens representing the top 10\% of the probability distribution. Higher top\_p values expand the sampling pool to include more tokens, thereby increasing the likelihood of selecting lower-probability tokens and consequently producing more random LLM outputs.

\textbf {Temperature:} Temperature influences the probability distribution of tokens after top\_p filtering \cite{sun2024source,karlsson2025reliable}. Higher temperature settings flatten the probability differences between tokens. To have increased scenario randomness, studies are more likely to select tokens with lower probabilities. When the temperature is set to 0, LLMs generate the same output consistently across repeated runs. Top\_p and temperature parameters work as mutually exclusive. Modifications should be made to only one parameter at a time, keeping the others constant \cite{openai2024chatcompletion}. 

We conducted an investigation across all three LLMs to identify the impact of temperature and top\_p on the quality of BDD scenario generation. As shown in Figure \ref{fig:BDD}, for each LLM in RQ4, we generated BDD scenarios using six different parameter combinations. Two top\_p values (0.5 and 1.0) crossed with three temperature settings (0, 0.5, and 1.0). We conducted three independent generation runs for each configuration to ensure reliability and used DeepSeek to evaluate the generated BDD scenarios on a 5-point scale. Final evaluation score calculated as the average across the three runs.

\begin{table}[t]
\centering
\caption{The impact of different model settings on LLM performance in generating BDD scenarios. \textbf{Bold} values indicate the best-performing temperature setting within each top\_p configuration. \textbf{Bold} and \underline{underlined} values show the best result for a model across all parameter settings. Grey \textbf{bold} and \underline{underlined} values represent the overall best score across all models and configurations.}
\label{tab:model_parameters}
\small
\begin{tabular}{llcccc}
\toprule
\textbf{Model} & \textbf{Top\_p} & \textbf{Metric} & \textbf{Temperature=0} & \textbf{Temperature=0.5} & \textbf{Temperature=1} \\
\midrule
\multirow{6}{*}{Claude 3} & \multirow{3}{*}{0.5} & Avg & \textbf{4.110} & 4.077 & 4.065 \\
& & MAD & 0.019 & 0.023 & 0.032 \\
& & Std. & 0.085 & 0.092 & 0.110 \\
\cmidrule(lr){2-6}
& \multirow{3}{*}{1.0} & Avg & \underline{\textbf{4.184}} & 4.076 & 4.088 \\
& & MAD & 0.225 & 0.037 & 0.039 \\
& & Std. & 0.307 & 0.119 & 0.121 \\
\midrule
\multirow{6}{*}{GPT-4} & \multirow{3}{*}{0.5} & Avg & 4.436 & \textbf{4.485}& 4.464 \\
& & MAD & 0.065 & 0.014 & 0.052 \\
& & Std. & 0.159 & 0.073 & 0.139 \\
\cmidrule(lr){2-6}
& \multirow{3}{*}{1.0} & Avg & \cellcolor[HTML]{EFEFEF}\underline{\textbf{4.626}} & 4.487 & 4.476 \\
& & MAD & 0.049 & 0.050 & 0.024 \\
& & Std. & 0.142 & 0.137 & 0.095 \\
\midrule
\multirow{6}{*}{Gemini} & \multirow{3}{*}{0.5} & Avg & \textbf{4.136} & 4.097 & 4.055 \\
& & MAD & 0.083 & 0.094 & 0.139 \\
& & Std. & 0.194 & 0.218 & 0.260 \\
\cmidrule(lr){2-6}
& \multirow{3}{*}{1.0} & Avg & \underline{\textbf{4.179}} & 4.129 & 4.099 \\
& & MAD & 0.036 & 0.124 & 0.117 \\
& & Std. & 0.116 & 0.266 & 0.250 \\
\bottomrule
\end{tabular}
\end{table}

\subsubsection {Experimental Results.}
The results in Table \ref{tab:model_parameters} reveal how these parameters influence the BDD scenario generation quality. We collected average scores, Mean Absolute Deviation (MAD) and standard deviation (Std.) across all parameter combinations. All three LLMs consistently achieved their best results with a temperature of 0 and a top\_p of 1.0 (shown as bold and underlined values in Table  \ref{tab:model_parameters}). Among them, GPT-4 achieved the highest performance (4.626 as highlighted in grey bold and underlined value in Table  \ref{tab:model_parameters}) at temperature 0 and top\_p 1.0. Claude 3 (4.184) and Gemini (4.179) demonstrated a similar pattern to GPT-4 at temperature=0 and top\_p=1.0. Once the temperature increased to 0.5 and 1, the quality of the BDD scenario generated by LLMs decreased. In almost all LLMs, the worst performance was achieved in temperature=1. The results confirmed that deterministic generates the best BDD scenarios compared to more random outputs. Also, the structured nature of Gherkin syntax helps with consistency rather than creative variation. All three LLMs performed better with top\_p=1.0 across all temperature settings. It proves that a full probability distribution improves BDD scenario quality.

\begin{center}
\label{sec:RQ3}
\begin{tcolorbox}[arc=1mm,width=1.0\columnwidth,
                  top=1mm,left=1mm,  right=1mm, bottom=1mm,
                  boxrule=.75pt]
\ding{43} \textbf{Key Finding} $\blacktriangleright$ Our experiments show that temperature=0 consistently produced the highest-quality BDD scenarios across all three LLMs (GPT-4, Claude 3, and Gemini). All models performed better with top\_p=1.0 across all temperature settings. The optimal configuration for BDD scenario generation is temperature=0 with top\_p=1.0, in which GPT-4 performed best. $\blacktriangleleft$
\end{tcolorbox}
\end{center}

\section {Discussion}\label{sec:discussion}
This section reflects on our findings and explores their implications for both research and industry practice.

\textbf{\textit{Practical Viability of LLM-based BDD Scenario Generation.}} Our research demonstrates that LLMs have reached a level of maturity where they can serve as a practical tool for generating BDD scenarios. The findings of RQ1 reveal that the evaluated LLMs, GPT-4, Gemini, and Claude 3, can generate high-quality BDD scenarios. This suggests that these models can accelerate existing BDD practices. However, the industry adoption of LLM-based BDD scenario generation faces several practical challenges beyond technical performance. Organisations implementing these processes must address concerns about cost-benefit analysis, integration with existing processes, and team training. We argue that successful adoption requires more than just deploying an LLM API endpoint; it requires integration into existing workflows and clear guidelines on when human review is essential.

Integrating LLM-based scenario generation into existing workflows must address tool compatibility to connect LLM generation with existing ecosystems (Jira, Confluence, Cucumber/SpecFlow, CI/CD pipelines). Process alignment concerns arise about whether automated generation should occur before, during, or after collaborative BDD discovery sessions (the “Three Amigos” approach \cite{agilealliance2023threeamigos}). Teams risk losing valuable stakeholder conversations if LLMs simply replace human collaboration. Accountability frameworks must clarify ownership and responsibility gaps, such as who owns generated scenarios and who is responsible when a generated scenario misses a critical defect that reaches production. To address these challenges, organisations should implement hybrid workflows in which LLMs produce initial scenario drafts for human refinement, maintaining collaborative value while improving efficiency. Gradual adoption, starting with low-risk features, builds team confidence before broader deployment. Clear responsibility frameworks in which human reviewers retain final accountability position LLMs as intelligent assistants rather than replacements.

\textbf{\textit{Matching Prompt Techniques to BDD Scenario Complexity.}} Variation in model performance across different prompting techniques (RQ2) introduces both opportunities and challenges for practitioners. Although GPT-4 demonstrated impressive performance with zero-shot prompting, Claude 3 showed improvement with chain-of-thought prompting, and Gemini achieved optimal results with few-shot examples. This finding challenges the idea that organisations can simply select \textit{the best} LLM and deploy it universally. Instead, teams must invest time in practical experimentation with different model-prompt combinations to identify what works best for their requirements. 

Organisations should consider maintaining a portfolio of prompting strategies tailored to different scenario types or complexity levels rather than assuming a single prompt template will serve all needs. Different BDD scenario characteristics may benefit from specific prompting techniques. Simple CRUD operations (Create, Read, Update, Delete) with straightforward logic may perform well with zero-shot prompting. Complex business rules with multiple conditional and interdependent logic may benefit from chain-of-thought prompting, where step-by-step reasoning helps decompose complex requirements. Integration scenarios with external systems or multi-step user workflows may require few-shot prompting with domain-specific examples to capture technical patterns and complete interaction sequences. During sprint planning, development teams can categorise user stories by these characteristics and select the appropriate prompting technique based on the scenario type. Future research should focus on the relationship between prompting techniques and BDD scenario characteristics, aiming to develop efficient and reliable BDD scenario generation.

\textbf{\textit{Role of Input Quality and Documentation Practices.}} Perhaps the most striking finding from RQ3 concerns the impact of input quality on scenario generation effectiveness. Our results demonstrate that detailed descriptions alone maintain high performance achieved with combined user stories and descriptions, while user stories alone result in low quality. This finding has profound implications for how organisations prepare their requirements documentation for LLM-based scenario generation. 

Traditional agile practices often use brief user stories, expecting details to emerge through conversation \cite{agilealliance2024userstory,atlassian2024userstories}. However, our research suggests that teams planning to leverage LLM-based BDD generation should invest more effort in creating comprehensive requirement descriptions. These descriptions should include specific acceptance criteria, edge cases, expected system behaviours, and technical constraints. This shift represents a change in documentation practices for many organisations and may require additional time investment during the requirements phase, though this investment is likely offset by the actual time savings in scenario generation.

Furthermore, the quality gap between description-only and user story-only inputs raises questions about the fundamental purpose of user stories in LLM-based workflows. If user stories prove insufficient for automated scenario generation, organisations might reconsider how they structure their requirement artifacts. One potential approach is to maintain user stories for stakeholder communication while developing parallel technical descriptions for use with LLMs. 

\textbf{\textit{Model Configuration and Creativity Trade-off.}}
The findings of RQ4 reveal a tension between \textit{consistency} and \textit{creativity} in BDD scenario generation. Our results demonstrate that temperature=0 with top\_p=1 consistently produces the highest-quality scenarios across all models, suggesting that deterministic generation outperforms more creative, randomised outputs for this specific task. However, this preference for consistency introduces questions about when, if ever, organisations should introduce randomness into scenario generation. Although our research focused on generating individual scenarios, real-world BDD practice often requires generating comprehensive test suites that cover multiple scenarios, edge cases, and failure cases. The consistency attribute excels at producing a single high-quality scenario but may yield redundant or overly similar scenarios when generating multiple test cases for the same requirement. Future research should investigate whether introducing controlled randomness (e.g., a temperature range of 0.3-0.5) during batch generation could improve scenario diversity without sacrificing individual scenario quality.

Industry practitioners must also consider the operational importance of model configuration choices. Deterministic generation (temperature=0) provides consistency benefits, such as the same input reliably producing the same output, facilitating debugging and quality assurance processes. However, this consistency means that incorrect prompts or input descriptions will consistently produce incorrect scenarios, potentially hiding quality issues until human review. Organisations might benefit from occasional sampling with higher temperature settings to identify edge cases or alternative scenario structures that deterministic generation might miss.

\textbf{\textit{LLM-based Evaluation and the Future of Quality Assessment.}} Our investigation of using LLMs as evaluators (RQ1) revealed that DeepSeek has the strongest correlation with human judgment. This finding aligns with recent research showing LLMs' potential as evaluators. Liu et al. \cite{liu2023g} demonstrated that GPT-4 achieves strong correlation with human judgments in Natural Language Generation (NLG) tasks, while Wang et al. \cite{wang2023chatgpt} found ChatGPT shows promise as an evaluator. However, the literature presents mixed evidence as Vykopal et al. \cite{vykopal2024disinformation} found that LLM evaluation capabilities vary significantly depending on task complexity and domain specificity.

Even our best-performing evaluator shows variance from human ratings across models. This indicates that automated evaluation cannot fully replace human expertise in assessing BDD scenario quality. Human evaluators bring domain knowledge, practical testing experience, and contextual judgment about edge cases that current LLMs cannot replicate \cite{haque2022semantic,shi2022evaluation}. The practical implication for industry is that organisations should adopt a hybrid evaluation approach. LLM-based evaluators like DeepSeek can efficiently screen large volumes of generated scenarios, identifying obviously poor outputs and flagging scenarios for human attention. Human experts can then focus their limited time on reviewing scenarios that automated evaluation marks as borderline or validating a sample. This approach will balance scalability with quality assurance.

\section{Threats to Validity}\label{sec:threats}
This section discusses potential threats to the validity of our research findings and the mitigation strategies employed to address them.
\subsection {Internal Validity}

\textbf {LLM Responses Inconsistency:} A primary threat to internal validity derives from the variability of LLMs, which can generate different responses to identical prompts across multiple executions. This randomness could compromise the reliability and reproducibility of our findings. We addressed this concern by setting the temperature parameter to 0 for all LLMs in RQ1, RQ2, and RQ3, which maximises the deterministic behaviour. Additionally, we conducted three independent evaluation runs for each experimental configuration and calculated average scores across these runs to minimise variance and strengthen result consistency.

\textbf {Limited Human Evaluation Sample:} We reduced the human evaluation set to 600 samples (200 per LLM and 50 per software product) from the 1,500 scenarios generated across all LLMs. We acknowledged that this subset of scenarios may not fully capture each model's performance spectrum. We had to maintain proportional distribution from the original dataset and keep the evaluation workload manageable for human experts.

\textbf {Evaluator Bias and Subjectivity:} Human evaluation of BDD scenario quality inherently involves subjective judgment, which can introduce bias despite our use of experienced QA professionals. We mitigated this threat by ensuring that each LLM's generated scenarios were independently evaluated by two separate evaluators, disagreements were identified and resolved through moderated consensus discussions, and all evaluators received standardised evaluation criteria on the 5-point rating scale. 

\subsection {External Validity}
\textbf {Domain-Specific Generalizability:} Our dataset was curated only from user stories from four software products developed by a single company. This limitation may restrict the generalizability of our findings to other software products such as embedded systems, mobile applications, or financial services, which may have different requirements. Although we acknowledge this limitation, we note that the company operates in 55 countries with various team structures, programming languages, and project scales, providing reasonable diversity within enterprise software contexts. Future research should validate these findings across additional products to establish broader applicability.

\textbf {Model Snapshot Validity:} LLM capabilities evolve rapidly, and our evaluation represents a snapshot of specific model versions (GPT-4, Gemini-1.5-Flash, Claude-3-Opus) at a particular point in time. Future model versions may demonstrate different performance characteristics. We clearly document the specific version of the models in Section \ref{sec:LLMs}. Our methodology and evaluation framework are designed to be reproducible with newer model versions, allowing future validation studies as models evolve.

\section{Conclusion and Future Work} \label{sec:coclusion}
This research addressed a gap in software testing by investigating the effectiveness of Large Language Models (LLMs) for automating Behaviour-Driven Development (BDD) scenario generation. Through an evaluation of three LLMs, GPT-4, Claude 3, and Gemini, using real-world data from 500 user stories across diverse software products, we provide evidence-based insights into the practical viability of LLM-based BDD generation. Our multidimensional evaluation framework, combining text-similarity metrics, semantic-similarity metrics, LLM-based evaluation, and human expert assessment, delivers key findings.
(1) GPT-4 and Claude 3 generate high-quality BDD scenarios, as evidenced by human evaluation scores. (2) LLM-based evaluators, mainly DeepSeek, demonstrate a stronger correlation with human judgment than traditional text similarity and semantic similarity metrics, offering a cost-effective alternative for large-scale assessment. (3) Optimal BDD scenarios generation requires model-specific prompting strategies: GPT-4 performs best with zero-shot prompting, Claude 3 benefits from chain-of-thought reasoning, and Gemini achieves optimal results with few-shot examples. (4) Input quality determines BDD generation effectiveness, with detailed requirement descriptions alone maintaining great performance while user stories alone result in low-quality scenarios. 

One of the primary contributions of this paper is the construction and release of the first publicly available dataset of 500 user stories with their corresponding BDD scenarios from real enterprise products. It establishes evaluation methodologies that balance automated efficiency with human expertise and provides direction for practitioners implementing LLM-based BDD automation. Our findings enable organisations to accelerate BDD scenario creation, reduce expertise barriers, and scale testing practices while maintaining the collaborative aspects that make BDD valuable. 

Although this research establishes a foundation for LLM-based BDD scenario generation, several directions deserve further investigation. Our work focused on individual happy-path scenarios, yet comprehensive BDD generation requires complete test suites covering edge cases and error conditions. Future research should explore multi-scenario generation strategies with comprehensive coverage, potentially using controlled randomness to balance quality. Our finding that DeepSeek demonstrates the strongest correlation with human judgment among LLM-based evaluators suggests promising directions for automated quality assessment research. Future work should investigate hybrid evaluation frameworks that combine multiple LLM evaluators and explore specialised evaluator models trained specifically for BDD scenario assessment, and develop evaluation approaches that provide actionable feedback.

\bibliographystyle{ACM-Reference-Format}
\bibliography{sample-base}
\end{document}